Complex Adaptive Systems, Publication 5
Cihan H. Dagli, Editor in Chief
Conference Organized by Missouri University of Science and Technology
2015-San Jose, CA

# Instantaneous Modelling and Reverse Engineering of Data-Consistent Prime Models in Seconds!

Michael A. Idowu

*SIMBIOS Centre, School of Science, Engineering and Technology, Abertay University, Dundee, DD1 1HG, United Kingdom*

## Abstract

A theoretical framework that supports automated construction of dynamic prime models purely from experimental time series data has been invented and developed, which can automatically generate (construct) data-driven models of any time series data in seconds. This has resulted in the formulation and formalisation of new reverse engineering and dynamic methods for automated systems modelling of complex systems, including complex biological, financial, control, and artificial neural network systems. The systems/model theory behind the invention has been formalised as a new, effective and robust system identification strategy complementary to process-based modelling. The proposed dynamic modelling and network inference solutions often involve tackling extremely difficult parameter estimation challenges, inferring unknown underlying network structures, and unsupervised formulation and construction of smart and intelligent ODE models of complex systems. In underdetermined conditions, i.e., cases of dealing with how best to instantaneously and rapidly construct data-consistent prime models of unknown (or well-studied) complex system from small-sized time series data, inference of unknown underlying network of interaction is more challenging. This article reports a robust step-by-step mathematical and computational analysis of the entire prime model construction process that determines a model from data in less than a minute.

## 1. Introduction

Complex systems are generally characterized by complex interactions of network components, each communicating with many other components in different ways, at different time intervals, and varying rates. Complex systems modelling research may be targeted at finding new complementary strategies that can improve existing knowledge about studied systems, project new line of thought and understanding of less-studied systems, or nominate new set of objectives for alternative views, hypotheses, experimental design and measurements. In line with these principal objectives, contemporary complex systems modelling, machine learning, and data mining research fundamentally often focus on model-based exploitation of time series measurements for instantaneous system identification and intelligent causal inference.

Here we introduce a new form of ODE-based discretization theory, apply knowledge of this to develop a new algorithm called TRM [1, 2, 3, 4, 5, 6], and demonstrate an application of TRM (causal inference) on test networks [2] and a real biological system using quantitative time series data of key kinases of the DDR pathway [6].

| **Nomenclature** | |
|---|---|
| ODE | ordinary differential equations |
| TRM | transposive regression (causal inference) method |
| DDR | DNA damage response |

*1.1. New model theory and discretization of continuous data*

In mathematical modelling, the concept of discretization applies to the partitioning of continuous data to acceptable smaller segments (in intervals) that can be used to represent the whole unpartioned (continuous) data, often carried out without loss of dynamic features. Here we reason that a good data discretization research should also aim at successful optimal recovery of original continuous data from discretized data. In [3] we evidenced successful complete recovery from artificial data discretized at small regular intervals under limited data condition, i.e. the total number of measured time points (states) equal to number of dependent (measurable) variables. The proposed discretization technique is based on a newly invented ordinary differential equations (ODE) theory formulated by the author [2, 4, 5, 6]. The proposed inference algorithm is discussed here in more detail with an illustrated analysis of presented here to show how the inference was performed.

*1.2. The interplay between interaction matrix, interaction network, and mathematical model*

Complex systems may be represented or aided by the use of network of interaction and network diagrams. This could be in the form of a graphical network model, interaction matrix, network of nodes in 2D, real physical network structure, or any other appropriate depiction for communicating and supporting plausible or functional causal explanation. For example, in systems biology interaction networks are used to describe biological phenomena and communication in genetic and biochemical networks [7, 8], further explain analysis results from clinical or bio-molecular data [9] and convey key concepts in personalized medicine [10]. Such representations may be linked to well-formulated process-based models that have captured and incorporated existing experts' knowledge extracted from the literature.

To understand the interplay between an interaction matrix, interaction network and model, the study of complex networks and use of network paradigm may be employed, seeking appropriate forms of network-based approach to describe dynamic systems, where the nodes in the network represent the (measurable) variables and the edges represent interactions between any pair of nodes. Crucially, interactions may have weighted edges where the weights relate to the strength of the interactions [7]. For example, gene networks are commonly represented by directed graphs where the nodes of the graph are genes and the directed edges are causal relationships between genes [7]. For example, in metabolic pathway modelling, it is important to formulate appropriate functions that describe the behaviour of the constituents of the system by identifying key components of interest with their symbolic names using (directional) arrows showing which components modulate the flows into, between, and out of components [8]. In such cases, the set of nodes is defined by the dependent variables; importantly, the edges and their weights, and so the network topology, must be inferred from the available data.

A major challenge confronting contemporary complex systems modelling is the problem of finding efficient and appropriate model construction methods to use. There is a necessary requirement to develop new, fast network inference techniques that may instantaneously solve parameter estimation tasks and construct network of interaction purely from discretized time series data. In this regard we propose a deterministic approach that is founded on matrix-based differential equations. The proposed method seeks to tackle system identification and parameter estimation challenges by adopting a matrix-based method of solving systems of linear (or alternatively, log-linear) differential equations.

### 1.3. Systems of ordinary differential equations

Conventional models that are based on ODE are regularly used to describe complex processes of a complex system [11]. The simplest form of ODE-based model is the deterministic system of linear differential equations (e.g. figure 1).

$$\begin{aligned}
\dot{X}_1 &= X_1 - 1.75*X_2 - 0.1685*X_4 + 1.331*X_7 + 0.6293*X_8 \\
\dot{X}_2 &= 0.6169*X_1 + X_2 + 0.1017*X_3 + 0.9019*X_5 - 0.4189*X_7 \\
\dot{X}_3 &= X_3 - 0.8671*X_2 + 0.5413*X_4 + 0.07005*X_6 + 0.8657*X_8 \\
\dot{X}_4 &= 0.6011*X_1 - 1.167*X_3 + X_4 - 0.6086*X_6 \\
\dot{X}_5 &= X_5 - 0.8985*X_2 - 1.223*X_6 + 0.3001*X_7 \\
\dot{X}_6 &= 1.778*X_4 - 0.5*X_3 + 2.227*X_5 + X_6 + 1.029*X_7 - 0.4381*X_8 \\
\dot{X}_7 &= 0.2908*X_2 - 0.6086*X_1 - 0.06921*X_5 - 1.343*X_6 + X_7 \\
\dot{X}_8 &= 0.5073*X_6 - 0.4336*X_3 - 0.7371*X_1 + X_8
\end{aligned}$$

Fig.1. Example of a simple ODE-based model

Alternatively, further interpretation and analysis of such simple models are easy to reformulated, particularly if intrinsically associated with matrix-based forms, e.g. the model in Fig.1 may be reformulated (Fig. 2).

| | | | | | | | | |
|---|---|---|---|---|---|---|---|---|
| $\dot{X}_1 =$ | 1*$X_1$ | -1.75*$X_1$ | 0*$X_1$ | -0.1685*$X_1$ | 0*$X_1$ | 0*$X_1$ | 1.331*$X_1$ | 0.6293*$X_1$ |
| $\dot{X}_2 =$ | 0.6169*$X_2$ | 1*$X_2$ | 0.1017*$X_2$ | 0*$X_2$ | 0.9019*$X_2$ | 0*$X_2$ | -0.4189*$X_2$ | 0*$X_2$ |
| $\dot{X}_3 =$ | 0*$X_3$ | -0.8671*$X_3$ | 1*$X_3$ | 0.5413*$X_3$ | 0*$X_3$ | 0.07005*$X_3$ | 0*$X_3$ | 0.8657*$X_3$ |
| $\dot{X}_4 =$ | 0.6011*$X_4$ | 0*$X_4$ | -1.167*$X_4$ | 1*$X_4$ | 0*$X_4$ | -0.6086*$X_4$ | 0*$X_4$ | 0*$X_4$ |
| $\dot{X}_5 =$ | 0*$X_5$ | -0.8985*$X_5$ | 0*$X_5$ | 0*$X_5$ | 1*$X_5$ | -1.223*$X_5$ | 0.3001*$X_5$ | 0*$X_5$ |
| $\dot{X}_6 =$ | 0*$X_6$ | 0*$X_6$ | -0.5*$X_6$ | 1.778*$X_6$ | 2.227*$X_6$ | 1*$X_6$ | 1.029*$X_6$ | -0.4381*$X_6$ |
| $\dot{X}_7 =$ | -0.6086*$X_7$ | 0.2908*$X_7$ | 0*$X_7$ | 0*$X_7$ | -0.06921*$X_7$ | -1.343*$X_7$ | 1*$X_7$ | 0*$X_7$ |
| $\dot{X}_8 =$ | -0.7371*$X_8$ | 0*$X_8$ | -0.4336*$X_8$ | 0*$X_8$ | 0*$X_8$ | 0.5073*$X_8$ | 0*$X_8$ | 1*$X_8$ |

Fig.2. Example of a reconfigured matrix-based ODE model (with heatmap)

Such a system is then described to above to be of linear differential equations with 8 dependent variables and an associated *Jacobian* transformation matrix *A* with the kinetic parameters

$$A = \begin{bmatrix}
1 & -1.75 & 0 & -0.1685 & 0 & 0 & 1.331 & 0.6293 \\
0.6169 & 1 & 0.1017 & 0 & 0.9019 & 0 & -0.4189 & 0 \\
0 & -0.8671 & 1 & 0.5413 & 0 & 0.07005 & 0 & 0.8657 \\
0.6011 & 0 & -1.167 & 1 & 0 & -0.6086 & 0 & 0 \\
0 & -0.8985 & 0 & 0 & 1 & -1.223 & 0.3001 & 0 \\
0 & 0 & -0.5 & 1.778 & 2.227 & 1 & 1.029 & -0.4381 \\
-0.6086 & 0.2908 & 0 & 0 & -0.06921 & -1.343 & 1 & 0 \\
-0.7371 & 0 & -0.4336 & 0 & 0 & 0.5073 & 0 & 1
\end{bmatrix}$$

Fig.3. Example of a simple ODE model

Furthermore, the transformation matrix *A* could be viewed as a representation of captures influxes and effluxes within and out of the interrelated components. For instance, the matrix A may be interpreted as illustrating complex interactions between the components (modes or dependent variables), each exerting positive and/or negative influences on other components of the network (Fig. 4).

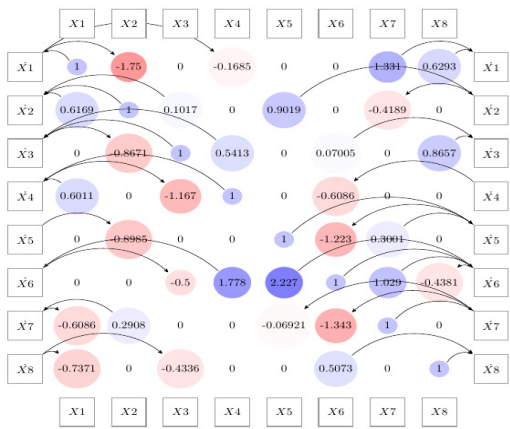

Fig.4. A transformation matrix as a mechanism of capturing and representing interactions (influxes and effluxes) within an interaction network

In general, prime models such as the above-mentioned may be employed to capture and represent systems of nonlinear equations whose general solution is equivalent to the system of linear (or log-linear) differential equations (Fig. 5).

$$\begin{bmatrix} \dot{X}_1 \\ \dot{X}_2 \\ \vdots \\ \dot{X}_n \end{bmatrix} = \begin{bmatrix} \frac{\partial X_1}{\partial X_1}X_1+ & \frac{\partial X_1}{\partial X_2}X_2 \cdots + & \frac{\partial X_1}{\partial X_m}X_m \\ \frac{\partial X_2}{\partial X_1}X_1+ & \frac{\partial X_2}{\partial X_2}X_2 \cdots + & \frac{\partial X_2}{\partial X_m}X_m \\ \vdots & \vdots & \vdots \\ \frac{\partial X_m}{\partial X_1}X_1+ & \frac{\partial X_m}{\partial X_2}X_2 \cdots + & \frac{\partial X_m}{\partial X_m}X_m \end{bmatrix}$$

$$\begin{matrix} X_1(t) \\ X_2(t) \\ \vdots \\ X_n(t) \end{matrix} = \begin{matrix} e^{\lambda_1 t}.[v_{1,1}].[p_1] + e^{\lambda_2 t}.[v_{1,2}].[p_2] + \cdots + e^{\lambda_n t}.[v_{1,m}].[p_m] \\ e^{\lambda_1 t}.[v_{2,1}].[p_1] + e^{\lambda_2 t}.[v_{2,2}].[p_2] + \cdots + e^{\lambda_n t}.[v_{2,m}].[p_m] \\ \vdots \\ e^{\lambda_1 t}.[v_{m,1}].[p_1] + e^{\lambda_2 t}.[v_{m,2}].[p_2] + \cdots + e^{\lambda_m t}.[v_{m,m}].[p_m] \end{matrix}$$

Fig.5. A model of linear differential equations (upper model); and an equivalent system of nonlinear equations based on eigenvalues and eigenvectors (lower image) [2]

The latter (i.e. nonlinear representation) adopts a formulation that may be based on eigenvalues and parameters or other forms of parameters and variables that must be consistent with the initial condition, e.g. linear combination of the eigenvectors, etc. As partially evidenced in [2], the author here concludes the implied solution must also be consistent with the *eigen-based* model (Fig. 6 below).

$$\begin{bmatrix} X_1(t) \\ X_2(t) \\ \vdots \\ X_m(t) \end{bmatrix} = e^{v\lambda v^{-1}t} \cdot \begin{bmatrix} v_{11} & v_{12} & \cdots & v_{1m} \\ v_{21} & v_{22} & \cdots & v_{2m} \\ \vdots & \vdots & \ddots & \vdots \\ v_{m1} & v_{m2} & \cdots & v_{mm} \end{bmatrix} \begin{bmatrix} p_1 \\ p_2 \\ \vdots \\ p_m \end{bmatrix}$$

Fig.6. A proposed eigen-based model which is consistent with the initial condition (IC) involving a decomposition of IC into linear combinations of eigenvectors.

## 2. Inverse Problems as Reverse Engineering Challenges

Assume that some evolutionary dynamics of a target complex system may be captured and recorded by measuring the amount, concentrations, or levels of key dependent (interacting) components over a sufficient time-period or certain (regular) time intervals. The unknown transformation matrix that defines the system characteristics or behavior, i.e. that have not been predetermined, must be inferred from the historical data in an efficient and consistent fashion. Ultimately the associated inverse problem must be completely solved to optimally identify the relevant and valid kinetics (approximated interactions and processes) that have occurred. In such scenarios where little or no information is supplied about the topology of the interaction network, prime models may be used to capture the unknown systems dynamics. The solution found may not be entirely unique, depending on the amount of data given, but it should adequate capture the observable features exhibited in the data. However, fast methods capable of producing plausible solutions are required, must be developed and engaged. For improved efficiency, there must be a devised non-iterative technique that can adequately guarantee a solution would be found under limited data condition and identify the unique solution in optimum time requirement and unlimited data conditions [2]. We assume a unique solution may exist if the data size (i.e. the total number of time points measured) exceeds the number of dependent variables.

*2.1. Inverse problems: systems of nonlinear equations*

Essentially, an inverse problem that is based on time series data may be formulated as a reverse engineering problem of finding a consistent matrix exponential from a system of nonlinear equations (figure 7).

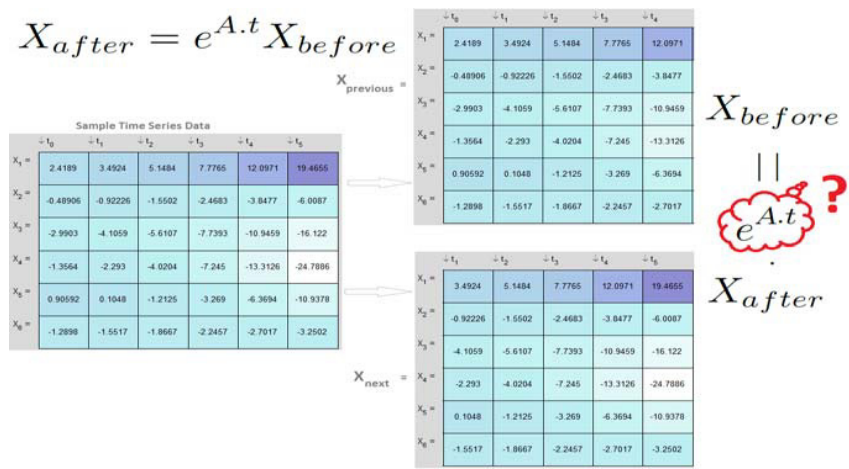

Fig.7. An elementary perspective on inverse problem definition of seeking to find optimal matrix exponential

Finding a matrix exponential is usually a difficult task; solving such without making *a priori* assumptions about the structure of the interaction network is much more challenging. The algorithm prescribed below introduces an efficient algorithm for solving inverse problems in seconds; the $X_{before}$ and $X_{after}$ vectors (defined and introduced above) represent the system states (i.e. array of vectors) before and after any transformations within the stated period, respectively.

## 2.2. A fast and efficient solution to inverse problems

Given an inverse problem where both the matrix exponential and transformation matrix must be found, the following algorithm may be used to find an appropriate solution. The transposive regression (TRM) method [2, 6], invented by the author, is presented in Fig. 8.

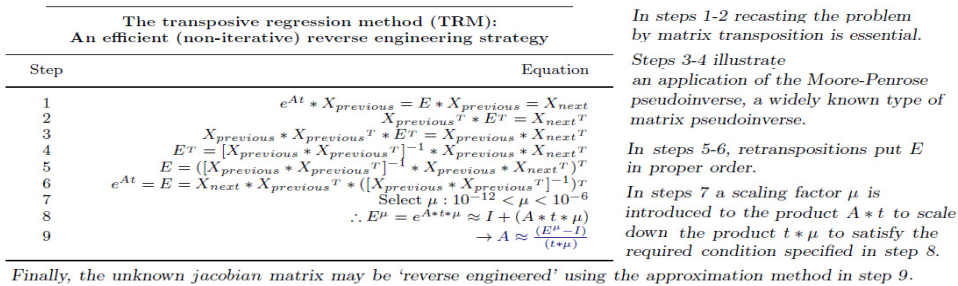

Fig.8. An invented TRM method for tackling system identification and parameter estimation challenges.

Using an appropriate discretization theory the derived transformation matrix must be assessed to verify that the actual historical data could be simulated. Our expectation is that a proposed inference method of strategy must first be simulating the original data that is supplied. This fundamental requirement should be satisfied and assessed in model construction before further simulation is performed or prediction made from the constructed model. For improved trust, reliability and evidencing performance any comparison made between the actual data and simulated data should minimize discrepancies during the assessment tests on solution methods.

## 3. Application of the proposed TRM method

### 3.1. Dynamic modelling of biological systems

Consider an application of the proposed reverse engineering method to DNA damage response (DDR) pathway modelling. A sample quantitative time series data of key kinases of the DDR pathway treated with an anti-cancer chemotherapy drug Doxorubicin is (measured and) recorded at (almost) regular time intervals (Fig. 9 – left image).

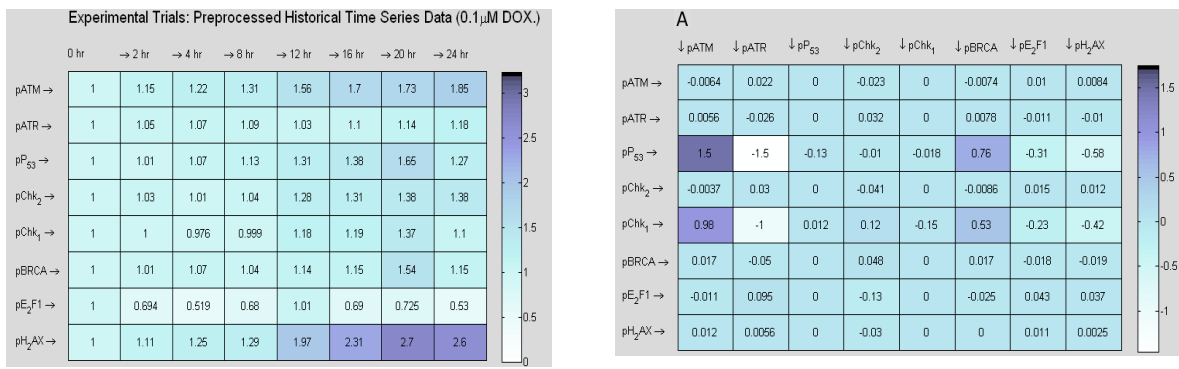

Fig.9. (a) A sample time series data (left) associated with the DNA damage response pathway; (b) the inferred *Jacobian* transformation matrix (right) consistent with the historical data [6]. The TRM inference method used to derive the transformation matrix ensures that the Jacobian-based model is consistent with the historical data as demonstrated in Fig. 10 (a).

The above data set typifies a limited data condition – the number of time points is quite small for effective dynamic modelling. The difficulty of the reverse engineering is even greater when the whole inference process must be purely based on the supplied data, i.e. in the absence of any additional constraints. The TRM method is then applied on the dataset measured at regular time intervals. Here in our case, this requires eliminating the 2hr-state vector (only) before analyzing the entire data. The transformation matrix (Fig. 9 – right image) is first derived using the TRM algorithm and should be verified (to ensure that it is capable of reproducing the historical data utilized) prior to simulating other missing (unmeasured, uncaptured and unrecorded) data within the recorded time periods. At the unknown interpolated time steps, the constructed model is then used to populate the actual dataset to produce a more continuous, far much richer quantity.

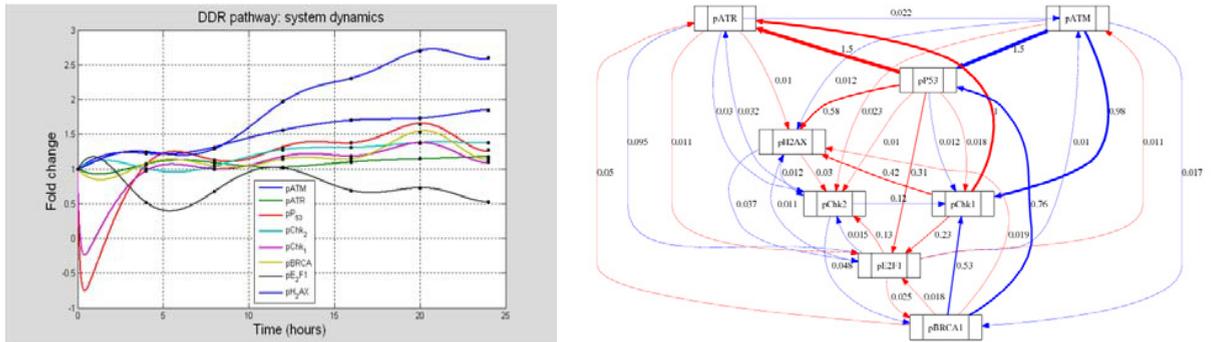

Fig.10. (a) An application of the TRM produces a non-discrepant comparison (left image) between actual (discretized experimental) data and predicted (simulated and interpolated) data ensuring data consistency; (b) Derived *in-silico* topological map representing revealed network of interaction extracted from the sample time series data. The derived topological map successfully evidences mild ATR activity, positive induction of ATM activity at the scale of DNA damage, a demonstration of pP53 S-15 to be the immediate downstream substrate of ATM and its positive regulation, signs of down regulation of E2F1 consistent with pATM-mediated E2F1 inhibition - characteristic that may further lead to cell cycle arrest [6].

With the inferred (Jacobian) matrix the constructed prime model of the biological system is able to simulate the complete dynamics of the system through continuous simulation; unknown dynamics are instantaneously generated to reconfirm the model outputs, verifying and ensuring little or no discrepancy between the actual and simulated (predicted) data as illustrated in Fig. 10 (left image). On successful identification of models that are consistent with data, a further system analysis of the optimal model would be required to further understand the revealed interactions extracted from the experimental data. Such an ODE-based causal inference strategy upgrades contemporary dynamic modelling and intelligent reverse engineering with data consistency and plausible extraction results, i.e. interaction network that may give adequate explanation to data in meaningful and understandable ways, e.g. Fig. 10 (right image).

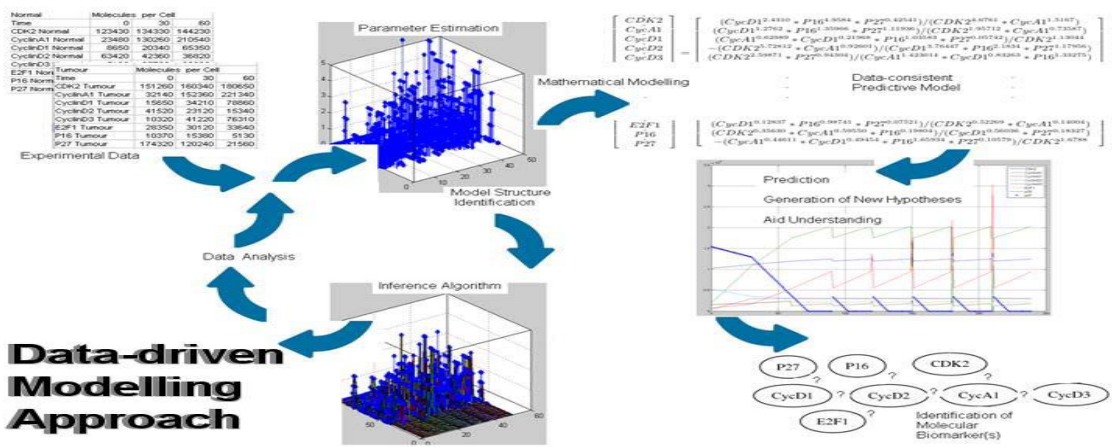

Fig.11. An overview of the proposed dynamic modelling method based on a direct utilization of the measured time series data and implementation of an optimization algorithm (e.g. the transposive regression method - TRM) for instantaneous system identification and network inference.

The proposed modelling strategies may be employed to analyse data of many complex systems, e.g. the proposed modelling and network inference technique is adopted to investigate and understand the role that certain key kinases play in the regulation of the cell cycle control system (Fig. 11). Beyond complex biological systems modelling, the TRM method may be implemented and in tackling other similar computational or mathematical challenges associated with complex adaptive systems, e.g. artificial neural networks (ANN), econometric modelling, inferential statistical analysis, advanced data analysis of complex dynamical systems, etc.

## 4. Conclusions

Dynamic processes in complex systems may be profiled by measuring system properties over time. One way of capturing and representing such complex processes (phenomena) is through an ODE model of the associated (measured) time series data. A novel computational method for reconstructing data-consistent ODE models in seconds has been invented by the author. The proposed technique could be used either as a purely data-driven modelling strategy or complementary approach to other existing process-based modelling methods. As demonstrated here, difficult challenges of instantaneous system identification (or automated parameter estimation) may be adequately tackled using the TRM algorithm. In underdetermined conditions, the solution produced might not be unique, however, the method always seeks a data consistent solution and avoid other candidate solutions that may not be plausible through the proposed discretization theory. In essence, our method demands utilization and reproducibility of original and additional historical data to improve intelligence and performance. Sample time series data are highly useful for instantaneous systems modelling, identification and improved design of optimization algorithms, inference methods and complex adaptive systems that require the support of theoretical network science.

## Acknowledgements

The author acknowledges the collaborative effort of Dr Hilal S. Khalil in performing the biological experiments that generated the sample DDR data and support received from Abertay university.